\documentclass[fleqn,usenatbib]{mnras}

\usepackage{newtxtext,newtxmath}

\usepackage[T1]{fontenc}
\usepackage{ae,aecompl}
\usepackage{multicol}

\usepackage{enumerate}
\usepackage[normalem]{ulem}
\usepackage[usenames]{color}

\usepackage{graphicx}	
\usepackage{ulem}

\newcommand{\changaMM}{{\sevensize MANGA}}
\newcommand{\changa}{C{\sevensize HA}NG{\sevensize A}}

\newcommand{\be}{\begin{eqnarray}}
\newcommand{\ee}{\end{eqnarray}}
\newcommand{\rsun}{\ensuremath{R_{\odot}}}
\newcommand{\msun}{\ensuremath{M_{\odot}}}

\newcommand{\unbeighty}{400}
\newcommand{\unbfull}{1500}
\newcommand{\thstart}{550}
\newcommand{\tinitial}{800}

\newcommand{\corr}[1]{#1}
\newcommand{\strikeout}[1]{}

\title[Envelope Ejection and Homologous Expansion in CEE]{Envelope Ejection and the Transition to Homologous Expansion in Common-Envelope Events}

\author[]{ Vinaya Valsan$^1$\thanks{ vvalsan@uwm.edu}, Sarah V. Borges$^{1}$, Logan Prust$^{1,2}$, Philip Chang$^1$  \\ $^1$ Department of Physics, University of Wisconsin-Milwaukee, 3135 North Maryland Avenue, Milwaukee, WI 53211, USA\\
$^2$ Kavli Institute for Theoretical Physics, University of California, Santa Barbara, CA 93106, USA}
\date{Accepted XXX. Received YYY; in original form ZZZ}

\pubyear{2022}

\newcommand{\vr}{\ensuremath{v_{\rm r}}}
\newcommand{\thomo}{\ensuremath{t_{\rm h}}}
\begin{document}
\label{firstpage}
\pagerange{\pageref{firstpage}--\pageref{lastpage}}
\maketitle

\begin{abstract}
We conduct a long-timescale ($5000\,$d) 3-D simulation of a common-envelope event with a $2\,\msun$ red giant and a $1\,\msun$ main sequence companion, using the moving-mesh hydrodynamic solver \changaMM. 
Starting with an orbital radius  of $52\,\rsun$, our binary shrinks to an orbital radius of $5\,\rsun$ in $200\,$d. We show that over a timescale of about $\unbfull\,$d, the envelope is completely ejected while $80$ per cent is ejected in about $\unbeighty\,$d.    
The complete ejection of the envelope is solely powered by the orbital energy of the binary, without the need for late-time reheating from recombination or jets.
Motivated by recent theoretical and observational results, we also find that the envelope enters a phase of homologous expansion about $\thstart \,\rm d$ after the start of our simulation. We also run a simplified 1-D model to  show that heating from the central binary in the envelope at late times does not influence the ejection. This homologous expansion of the envelope would likely simplify calculations of the observational implications such as light curves.
\end{abstract}


\section{Introduction}
Common-envelope evolution (CEE; \citealt{Paczynski:1976}) is believed to be responsible for the production of many close binary systems such as  X-ray binaries, binary neutron stars, binary black holes, and white dwarf binaries including cataclysmic variables \citep[for a review, see][]{Ivanova:2013}.  The physics and astrophysics of these systems have been a subject of continuous study for the last $50$\,yr. However, the complex physics of this process, which includes gravity, hydrodynamics, nuclear burning, recombination, and radiation, has precluded much analytic progress.  In the last decade, advances in computing and algorithms have made high-resolution, long-timescale simulations of CEE possible.  

These long-timescale simulations have begun to unravel the relevant physics of CEE and its astrophysical impact.  
In particular, studies carried out by several groups on low-mass binaries such as an asymptotic giant branch (AGB) primary with a main-sequence (MS)  companion star \citep{Sand:2020, Chamandy:2020, Ondratschek:2022} and a red-giant branch (RGB) primary with an MS star \citep{Iaconi:2019}  have demonstrated that the envelope is completely ejected on sufficiently long timescales (years to decades).  
These groups differ in their conclusions of what physics is important.  For instance, \cite{Iaconi:2019}, \cite{Sand:2020}, and  \cite{Ondratschek:2022} argue that recombination energy is crucial in envelope ejection, but \cite{Chamandy:2020} argue otherwise.
This difference is likely due to the limited number of cases studied and future studies will likely bring this physics into sharper focus.

In addition, recent work by \cite{Iaconi:2019} argues that the expansion of the envelope leads to homologous expansion on long timescales. 
This should not come as a surprise because any expansion leads to homologous expansion so long as the trajectories remain ballistic. 
However, if the envelope expands homologously, this helps to simplify the theory of CEE and 
would provide an easier means to compute observables.

Recent observations have also hinted at the presence of homologously expanding ejecta in CEE and stellar merger events. 
For instance, the properties of observed CO emission in V4332\, Sgr is well reproduced with the homologous expansion model \citep{Kaminski:2018}.  Additionally, \cite{Kaminski:2020} showed that the observed properties of the molecular remnant of Nova 1670 (CK Vulpeculae) are satisfactorily reproduced by linear velocity fields.

In this paper, we study the physics of the ejection of the envelope using long-timescale simulations.  
Starting with our recent work \citep{Logan:2019, Logan:2020}, we optimize our numerical techniques to allow for an order of magnitude increase in integration time. 
We show that for the case of a $2\,\msun$ RGB and a $1\,\msun$ MS companion, we achieve complete envelope ejection in $\unbfull\,$d. We also show that the envelope enters a homologous phase early on (about $\thstart\,$d) and that the morphology of the ejected material is roughly spherical.

The paper is organized as follows.  In Section~\ref{sec:setup}, we discuss the numerical setup of our calculations.  
We follow \cite{Logan:2019} and \cite{Logan:2020}, but describe a significant improvement in how we generate initial conditions, giving substantial speedups.  
In Section~\ref{sec:results}, we discuss the results of these simulations and show complete envelope ejection after $\unbfull\,$d. We also demonstrate that the envelope enters a homologous phase early on and that it can be approximated as spherical.  
Motivated by these results, we develop a 1-D numerical model in Section~\ref{sec:1d} and compare these simplified calculations with the full 3-D calculation.  
In Section~\ref{sec:discussion}, we discuss the major theoretical and observational implications of our results and close in Section~\ref{sec:conclusions}.

\section{Numerical Setup} \label{sec:setup}
We use the moving-mesh hydrodynamic solver for \changa, which we call \changaMM~  \citep{Chang:2017, Logan:2019}, to study CEE.  
\changaMM\ is a moving-mesh hydrodynamic simulation code based on the algorithms described by \cite{Springel:2010}. 
A detailed description of \changaMM\ is presented by \cite{Chang:2017}.  Improvements such as the multiple time-stepping algorithms and integration of stellar equations of state are presented by \cite{Logan:2019}.  
A discussion of its use for simulations of main sequence tidal disruption events is presented by \citet{Alexandra:2021}.  
Finally, recent code improvements are discussed by \citet{Chang:2020a} for radiation hydrodynamics, \citet{Chang2020b} for general relativistic hydrodynamics, \citet{Logan:2020} for moving-boundary conditions, and Prust \& Chang (in preparation) for magnetohydrodynamics. We refer the interested reader to this literature for a detailed description of \changaMM.

\subsection{Initial Conditions}
We use the same procedure as \citet{Logan:2019} to construct initial conditions.  We evolve a $2\,\msun$ star with MESA \citep{Paxton:2011,Paxton:2013,Paxton:2015,Paxton:2018,Paxton:2019,Jermyn:2022} from the pre-main sequence to the red giant phase and stop when it reaches a radius of around $52\,\rsun$.  
As done by \citet{Logan:2019}, we construct a star of mass $2\,\msun$, whose entropy profile matches that of the original star. The core of the newly-constructed star is modelled as a dark matter particle with gravitational softening for computational expediency.  We then map the radial profile of density and temperature to an unstructured particle (mesh-generating point) mesh. The simulation consists of 430K mesh-generating points, of which 80K model the star. 
The companion star, which is also modelled as a softened dark matter particle, is then placed at the surface of the red giant.

\corr{We should note that placing the companion on the surface of the red giant as an initial condition is unrealistic for a couple of reasons. First, the red giant model used at the beginning of the simulation is only in hydrostatic equilibrium in isolation and thus does not account for the effects of the companion. Second, a more realistic scenario would involve the binary system evolving, allowing the red giant to slowly expand on a nuclear timescale until it fills its Roche lobe. At this point, it undergoes unstable mass transfer, driving the system into CEE.  However, this more realistic scenario is not easily realizable in numerical simulation as it involves the slow evolution of the red giant in the thermal timescale and would demand a realistic treatment of nuclear burning and radiation.  The computational costs of such a simulation would be prohibitive.  Thus, we have simplified the initial conditions to the ones stated and anticipate that for the long-term evolution of the CEE event, the initial conditions do not play a large role.}

In this paper, we have made a number of modifications in an effort to reduce the computational cost. 
First, we use an adiabatic equation of state ($\gamma=5/3$) instead of a MESA equation of state. 
\corr{The simulation using the MESA equation of state is significantly more computationally expensive than the adiabatic case. 
The primary advantage of using the MESA equation of state is that it encodes additional information in regard to recombination energy, which may be important in ejecting the envelope for low-mass systems.  
Nevertheless, as we will demonstrate below, orbital energy alone achieves complete envelope ejection without the need for recombination energy.}
We also improved the grid generation for the tenuous atmosphere surrounding the stars. 
In particular, we increase the spacing between grid points exponentially outside the stars up to the final coarsest resolution instead of the power-law increase used previously by \citet{Logan:2019}.  
This reduces the number of nearest-neighbour searches for mesh-generating points near the boundary of the star and atmosphere. 
We find that this improves the performance of the code by around a factor of 2.  

These reductions in computational costs and improved computing power allow us to run our simulation for $5000$\,d, which is longer than our previous simulations \citep{Logan:2019} by a factor of around $20$.  It is also in line with other recent long-timescale simulations \citep{Iaconi:2019, Sand:2020}. 
\section{Results}\label{sec:results}
\begin{figure*}
\includegraphics[width=0.47\textwidth]{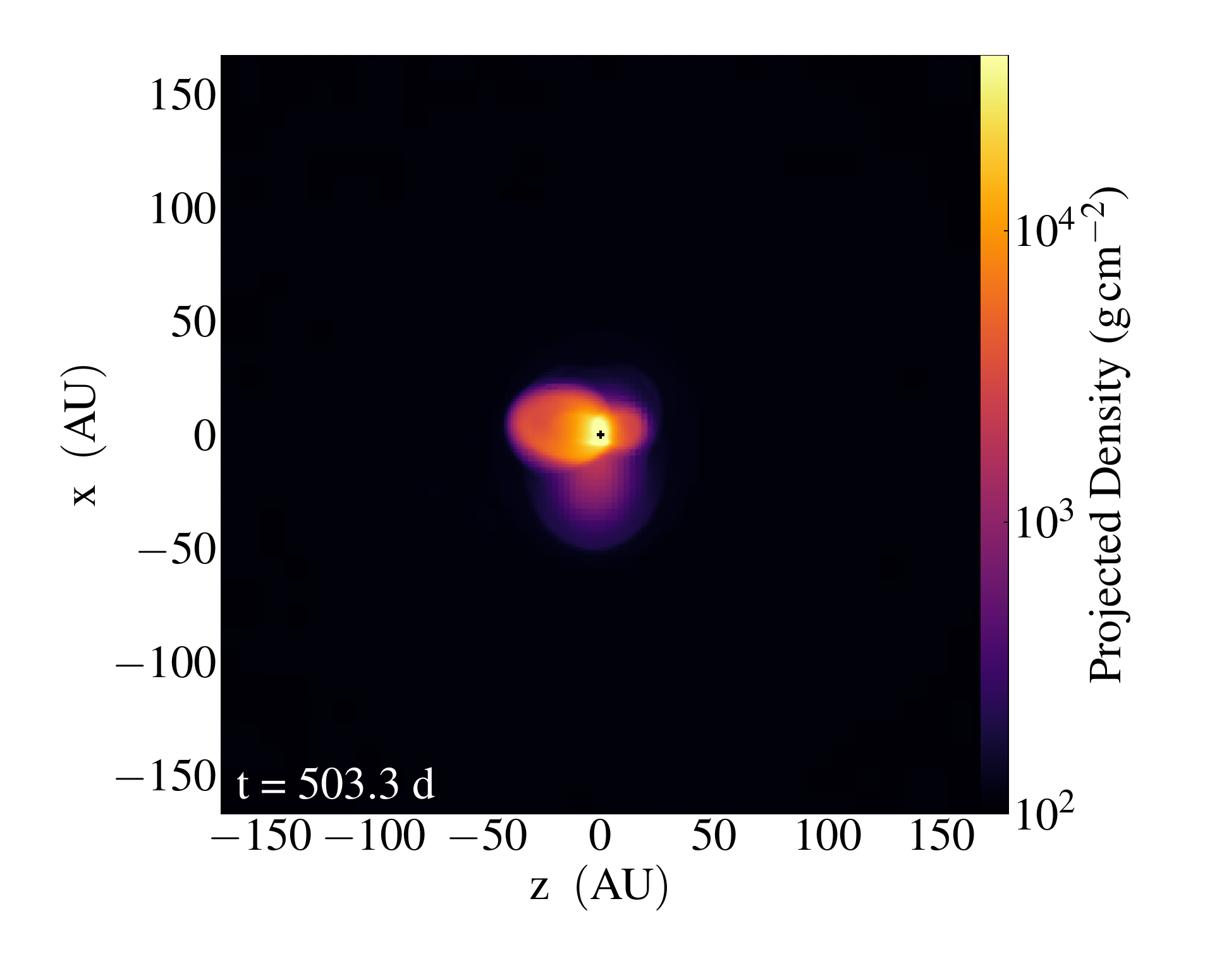} 
\includegraphics[width=0.47\textwidth]{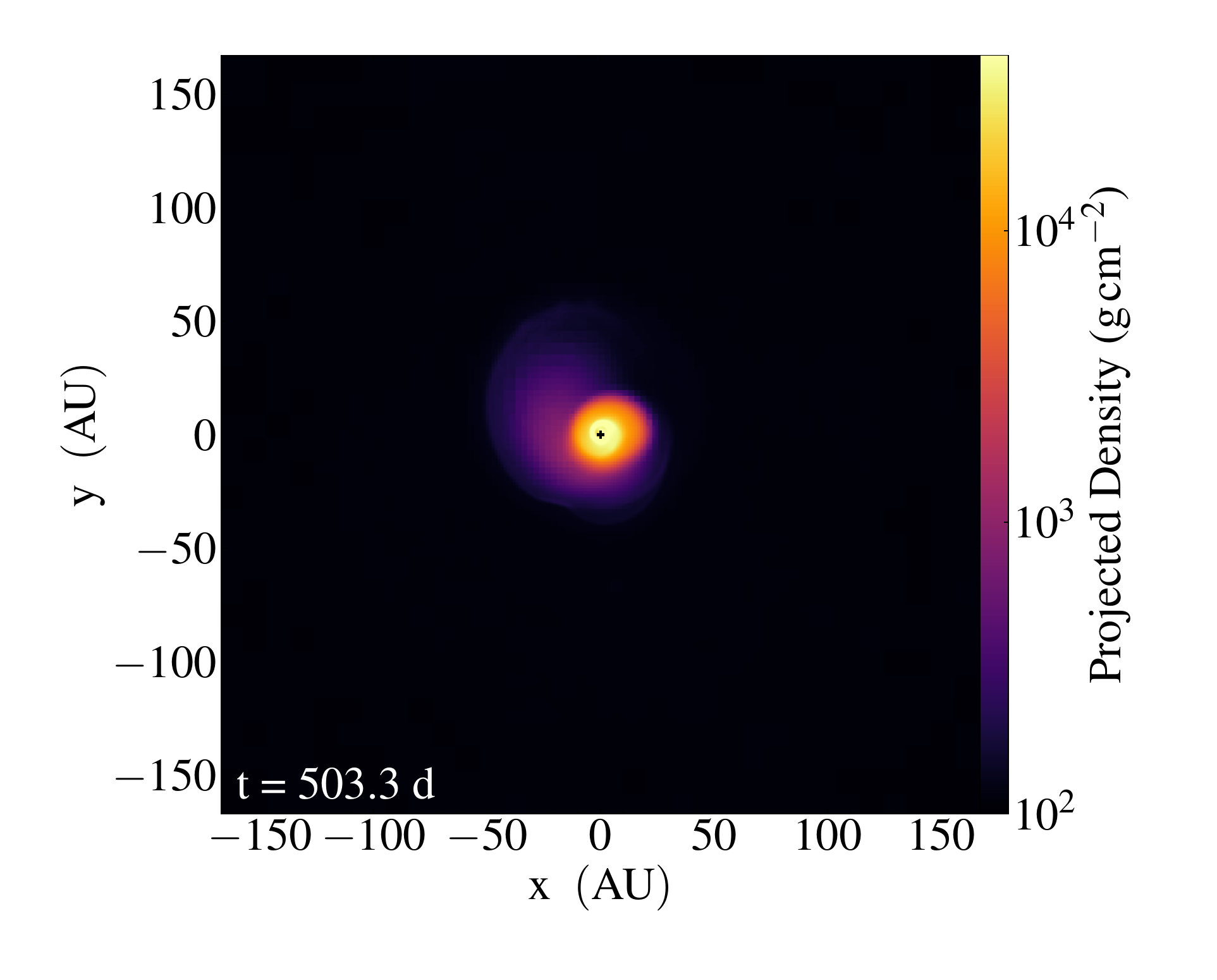}\\
\includegraphics[width=0.47\textwidth]{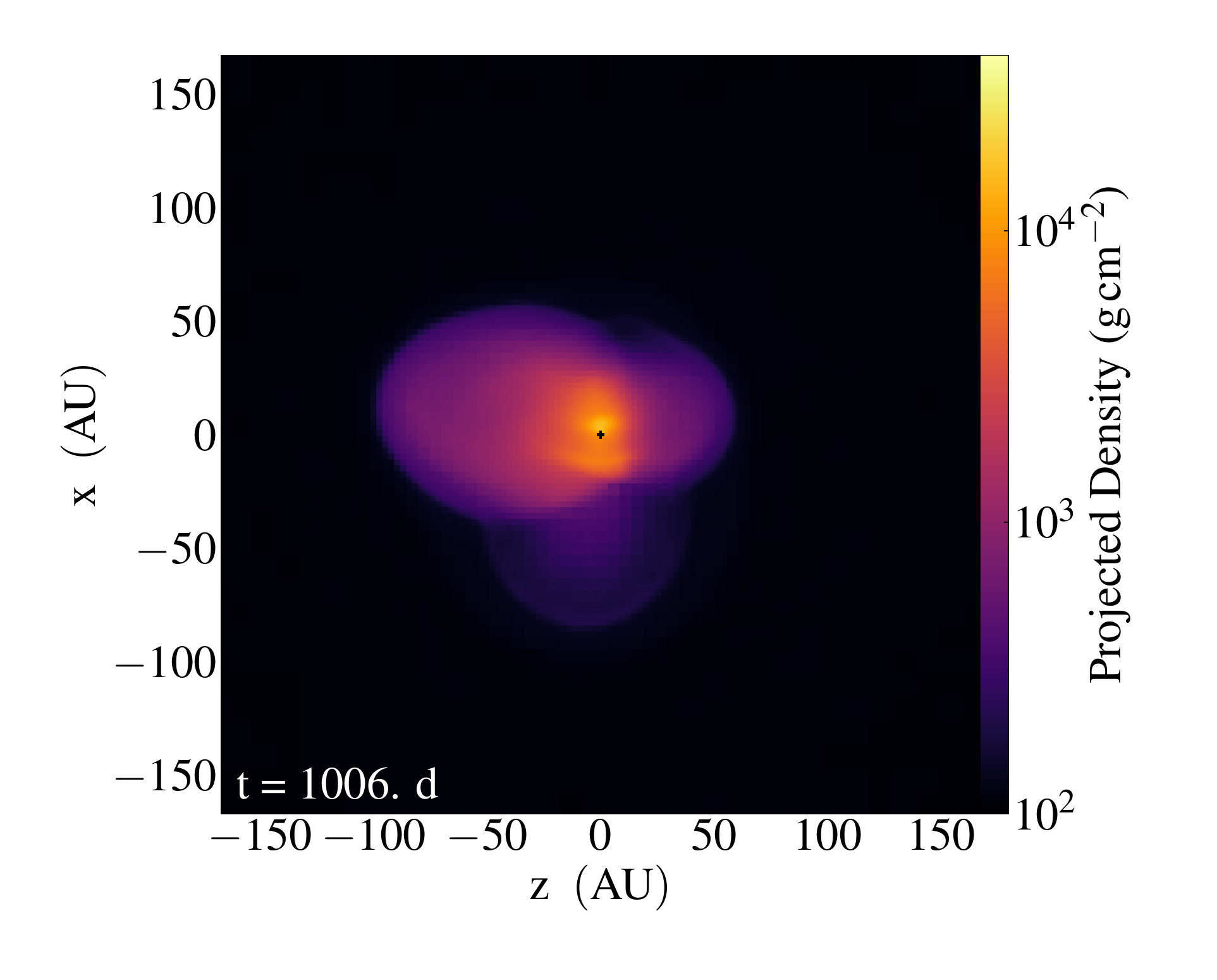} 
\includegraphics[width=0.47\textwidth]{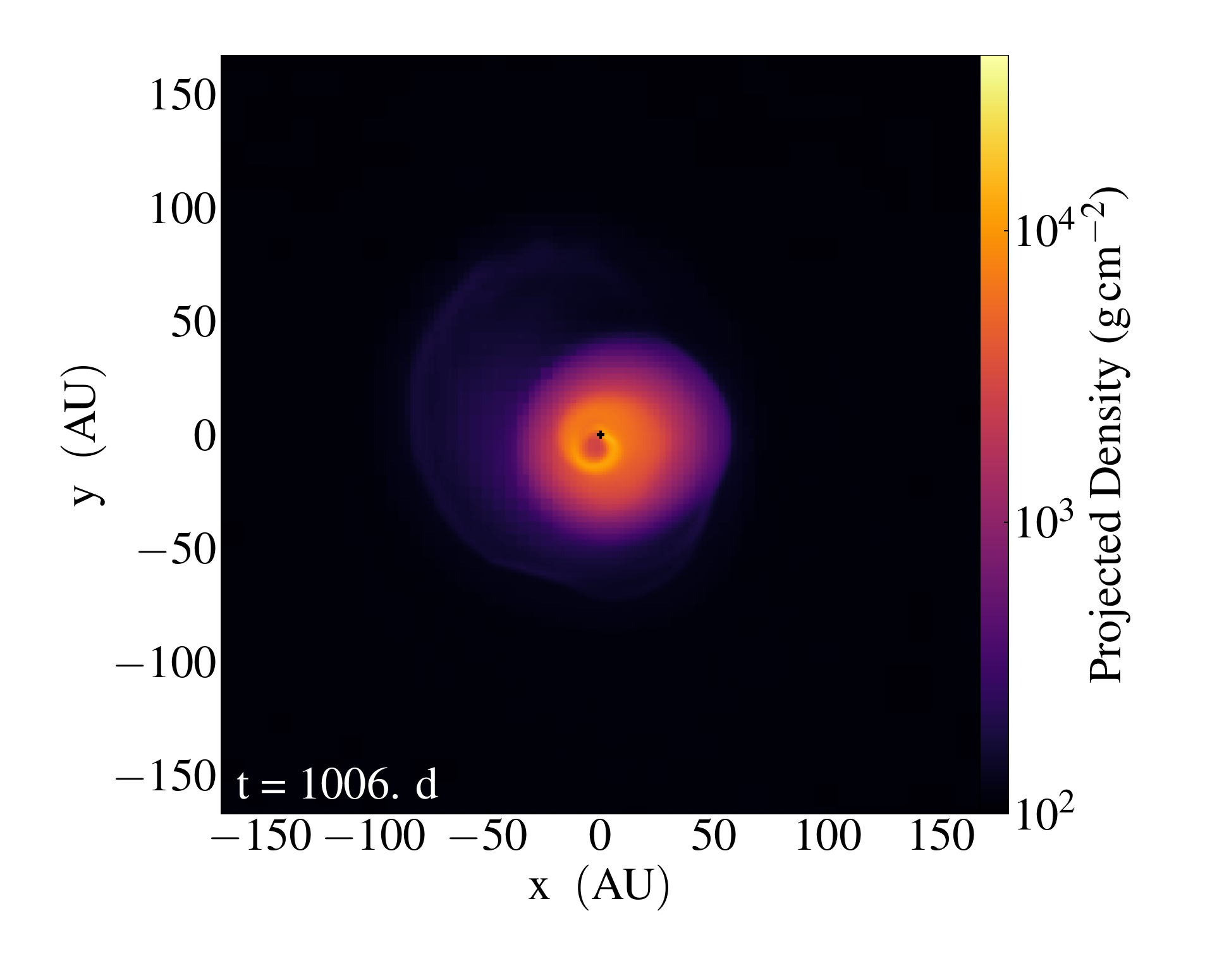}\\
\includegraphics[width=0.47\textwidth]{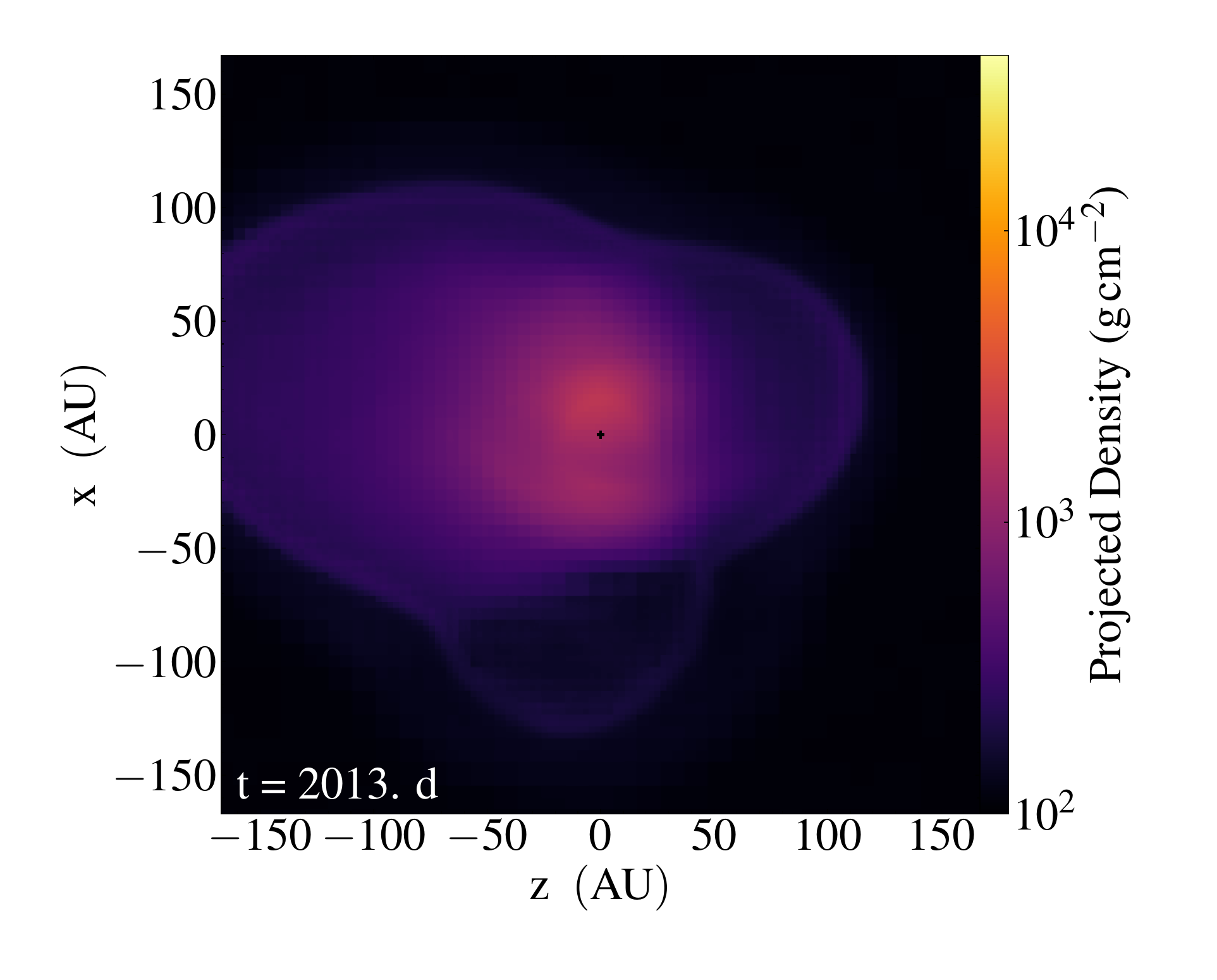} 
\includegraphics[width=0.47\textwidth]{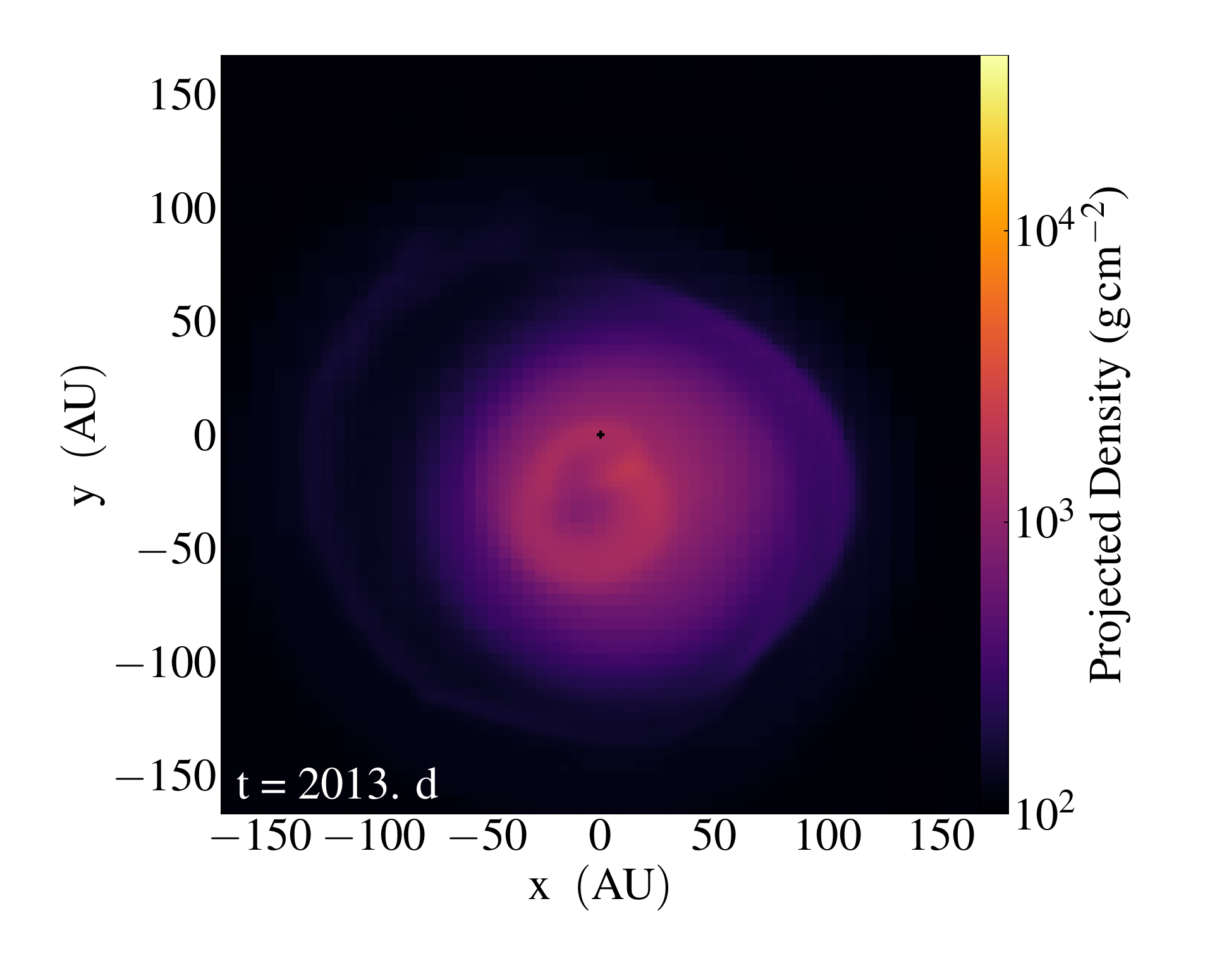}
\caption{Projection of density on to the $x-z$ plane (left panel) and $x-y$ plane (right panel) at different time slices ($503\,$d, $1006\,$d, $2013\,$d)
from the simulation. The `$+$' sign marks the centre of mass of the system.}
\label{fig:densityproj}
\end{figure*}

\begin{figure}
\includegraphics[width=0.47\textwidth]{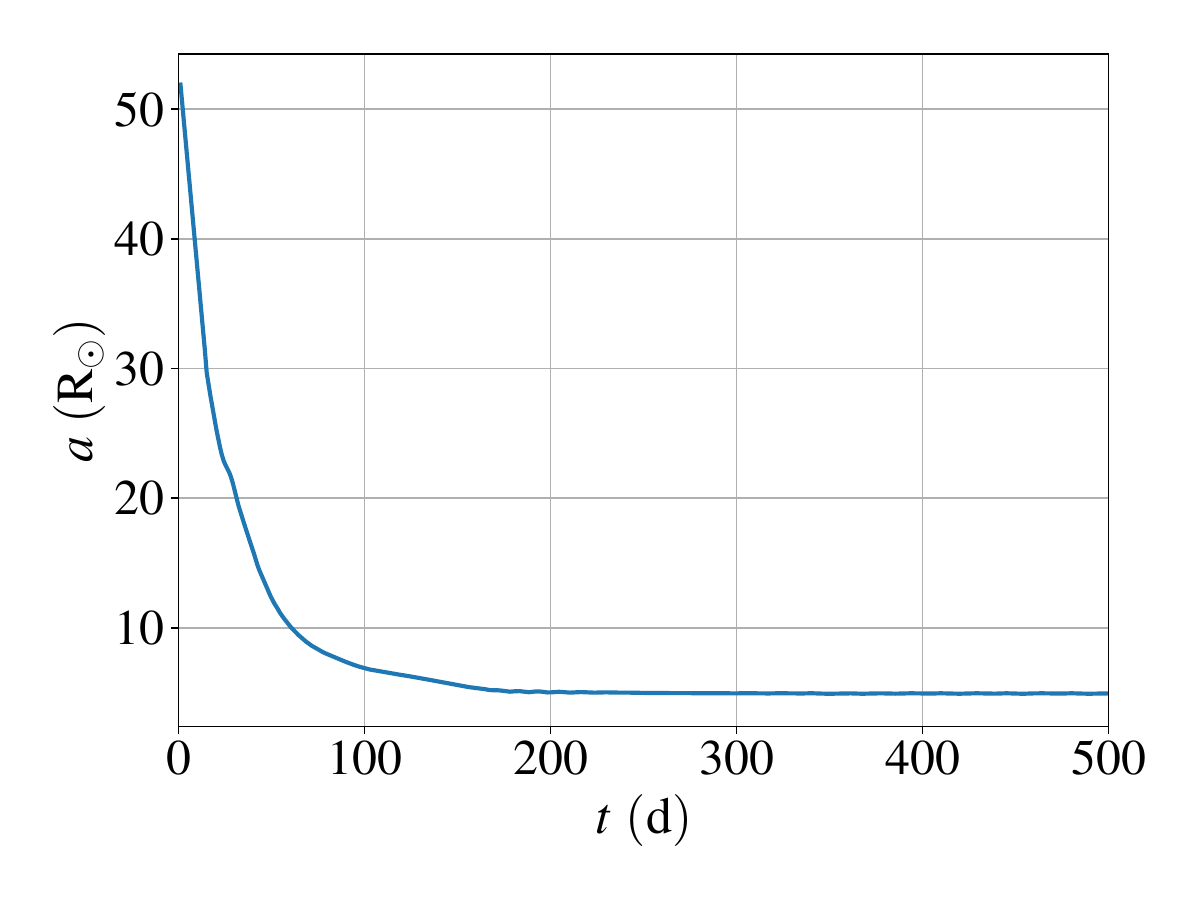}
\caption{Smoothed separation between the centre of mass of the red giant and the main-sequence star as a function of $t$.  
There is an initial rapid plunge of the two centres toward each other, but this plateaus to about $ 5\,\rsun$ after $200\,$d. 
}
\label{fig:sep}
\end{figure}

We show a series of projected density plots in Fig.\ref{fig:densityproj}, projected along the axis of the orbital plane (z-axis, right column) and projected along a direction in the orbital plane (x-axis, left column) at $503\,$d, $1006\,$d and $2013\,$d.
The `+' sign marks the centre of mass of the system. Initially, these projections demonstrate that the ejected matter is axisymmetric, but not spherical.  
The overall shape remains fairly constant, though it does become more spherical as it evolves.  
This constancy of the overall shape and evolution to greater spherical symmetry will be important in our discussion of homologous expansion and the spherical approximation.

In Fig.~\ref{fig:sep}, we plot the orbital separation, $a$, between the centres of the stars.  
Each  $a$ in the plot is taken to be the average of $a$ in a time interval of 14 d. 
In the beginning, the two centres plunge toward one another in a period of rapid orbital decay.  
This starts to slow down and at $200$\,d, the orbital decay plateaus to an orbital separation of $a\approx 5\,\rsun$, which is similar to our previous result of $a \approx 3.6\,\rsun$  \citep{Logan:2019}. 

\begin{figure}
\includegraphics[width=0.47\textwidth]{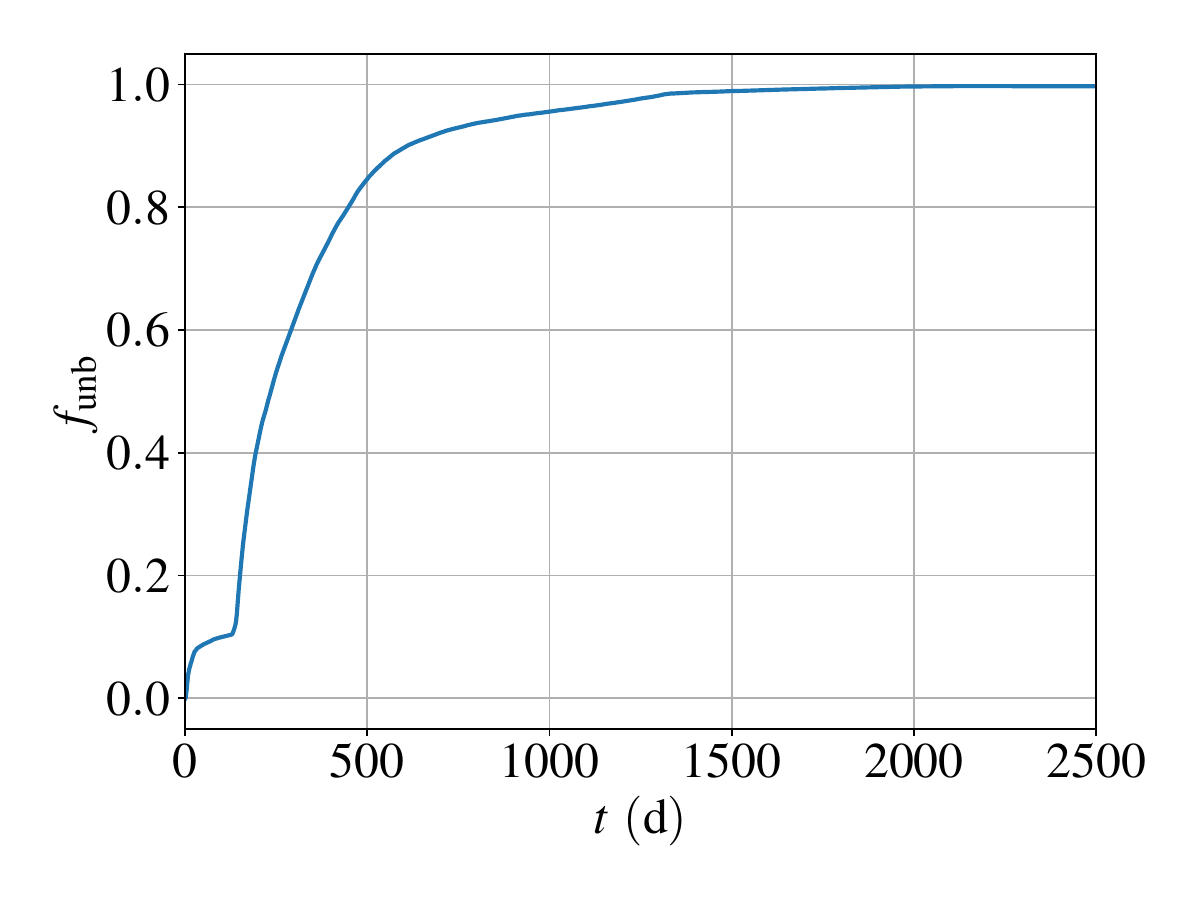}
\caption{Fraction of unbound mass, $f_{\rm unb}$, as a function of time, $t$.  We only consider the mechanical energy in this case.} \label{fig:unbound}

\end{figure}

For each $\rm i$-th particle/mesh-generating point in the system, the total mechanical energy is defined as
\be
    E_{\rm mech,i} = m_{\rm i}\left(\frac{1}{2}v_{\rm i}^2+\phi_{\rm i} \right),
\ee
where $m_{\rm i}$ is the total mass of the mesh-generating point, $v_{\rm i}$ is the fluid velocity of the mesh-generating point relative to the bound centre of mass of the system, and $\phi_{\rm i}$ is the gravitational potential. 
Particles with $E_{\rm mech,i} < 0$ are bound to the binary while those with $E_{\rm mech,i} > 0$ are unbound.  
As discussed by \citet{Logan:2019}, we must carefully define the velocities relative to that of the centre of mass of the bound material.  
This involves an iterative computation to find the bound mass and the centre of mass velocity. 
The unbound mass fraction is then defined as the fractional mass of the material with positive total energy.  

In Fig.~\ref{fig:unbound}
we plot the unbound mass fraction, $f_{\rm unb}$, as a function of time. 
Nearly all gas from the red giant is unbound after $\unbfull\,$d and over $80$ per cent at $\unbeighty\,$d. 
At $250\,$d, the fraction of mass that is unbound compared to our previous result \citep{Logan:2019} is substantially larger (40 per cent vs. 10 per cent), considering just the mechanical energy.  
However, the equation of state is different between the two simulations (ideal gas vs MESA).  
In addition, envelope ejection occurs in the absence of additional late-time energy injection via hydrogen recombination and/or jets.  
Other work has also recently demonstrated complete envelope ejection on a time scale of about $1000\,$d, but these results can rely on additional late-time energy injection.



\subsection{Homologous Expansion}

Recently, \citet{Iaconi:2019} showed in their long-timescale CEE simulations that the envelope ejection follows a homologous expansion approximation.  
In their work, they simulate the evolution of a $ 0.88\,\msun, 83\,\rsun$ RGB and a $0.6\,\msun$ companion star and follow the system for about $15$\,yr. 
They show that the external layers of the envelope become homologous as soon as they are ejected, but that it takes about $14$\,yr for the bulk of the unbound gas to enter homologous expansion. 
Motivated by this result, we investigate the onset of homologous expansion in our simulations.

To begin, we recall that the distinguishing characteristic of homologous expansion is that the velocity follows a radial profile $v\propto r$.  In essence, this means that fluid elements are on ballistic trajectories with little or no interaction between fluid elements or external forces. As such, the radial position of a fluid element can then be written as 
\be\label{eq:r(t)}
r(t) = \vr \thomo,
\ee
where $\vr$ is the radial velocity of the fluid element and $\thomo$ is the homologous expansion time.  We note that while the formalism of homologous expansion is relatively simple and discussed widely in the literature, we discuss it here again to define it in terms of computational quantities like time, which is defined as zero at the beginning of a simulation and has no relation to the zero time in homologous expansion.  Thus, we define $\thomo = t - t_0$, where $t$ is the time since the start of the simulation and $t_0$ is some fitted time that defines the $t=0$ point of homologous expansion.  Indeed, $\thomo$ is mapped exactly to the time in most discussions of homologous expansion.
We can rewrite equation (\ref{eq:r(t)}) as
\be
\label{eq:vrad}
    \vr = \frac{r}{t-t_0} = \frac{r}{\thomo},
\ee
and by differentiating and integrating the above equation, we can write the position of a fluid particle at time $t$ with respect to an initial time $t_{\rm i}$ as 
\be\label{eq:rt}
r(t) = r_{\rm i}\frac{t-t_0}{t_{\rm i}-t_0} = r_{\rm i} \frac{\thomo}{t_{\rm {h,i}}},
\ee
where $r_{\rm i}$ is the radial position at a time $t_{\rm i} > t_0$.  
From equation (\ref{eq:rt}), we now define a scaled radius with respect to the initial time given the current radius for any fluid element in the simulation
\be\label{eq:rs}
    r_{\rm s}(r,t) = r\frac{t_{\rm h,i}}{\thomo} = r\frac{t_{\rm i}-t_0}{t-t_0}.
\ee
In other words, $r_{\rm s}$ maps the position of a fluid element, $r$, at a particular time $t$ to the position of a fluid element at the initial time $t_{\rm i}$. Having defined the scaled radius, $r_{\rm s}$, we also define the scaled density and velocity as
\be\label{eq:rhos}
    \rho_{\rm s}(r,t) &=& \rho_{\rm i}(r_{\rm s}(r,t)) \left ( \frac{\thomo}{t_{\rm h,i}} \right )^3\\
    v_{\rm r,s}(r,t) &=& v_{\rm r,i}(r_{\rm s}(r,t)).\label{eq:vs}
\ee

The simple rescaling given in equations (\ref{eq:rhos}) and (\ref{eq:vs}) is insufficient to describe the entire system.  While it works for the expanding envelope, it does not describe the tenuous atmosphere.  Toward that end, we define the radius of the envelope, $R(t)$, which is the position of the outer boundary of the homologously expanding region inside  which equations (\ref{eq:rs}), (\ref{eq:rhos}) and (\ref{eq:vs}) are valid.  We define a dimensionless radius $\eta$ as
\be
\eta = \frac{r}{R},
\ee
so that $\eta=0 $ at the centre of the CE and $\eta=1$ at the envelope's outer boundary.
From equation (\ref{eq:rt}), we can estimate $R$ for any given time with respect to the initial time $t_{\rm i}$: $R = R_{\rm i} ({\thomo}/{t_{\rm h,i}})$. 
We can then redefine scaled density, $\rho_{\rm s}$, and scaled velocity, $v_{\rm s}$, for regions inside and outside of the envelope.  This gives
\be
  \rho_{\rm s} &=&
    \begin{cases}
        \rho_{\rm i}(r_{\rm s}(r,t)) \left(\frac{\thomo}{t_{\rm h,i}} \right) ^3 & \text{if $\eta\leq 1$}\\
      \rho_{\rm b} & \text{if $\eta$ > 1} 
    \end{cases}     \nonumber  \\
    v_{\rm r,s} &=&
    \begin{cases}
        v_{\rm r,i}(r_{\rm s}(r,t))  & \text{if $\eta\leq 1$}\\
      0 & \text{if $\eta$ > 1} 
    \end{cases}  \label{eq:scalings}
\ee

\begin{figure}
    \includegraphics[width=0.47\textwidth]{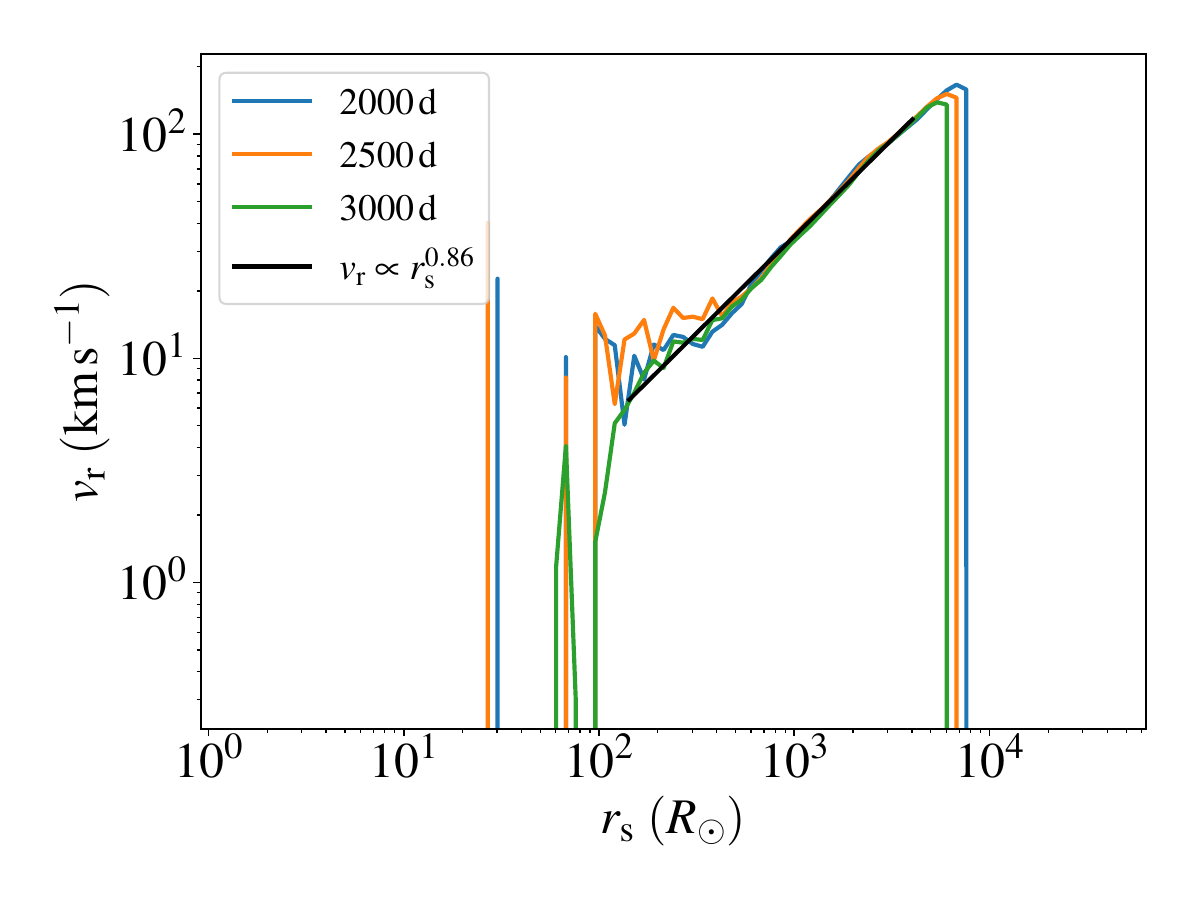}\\
    \includegraphics[width=0.47\textwidth]{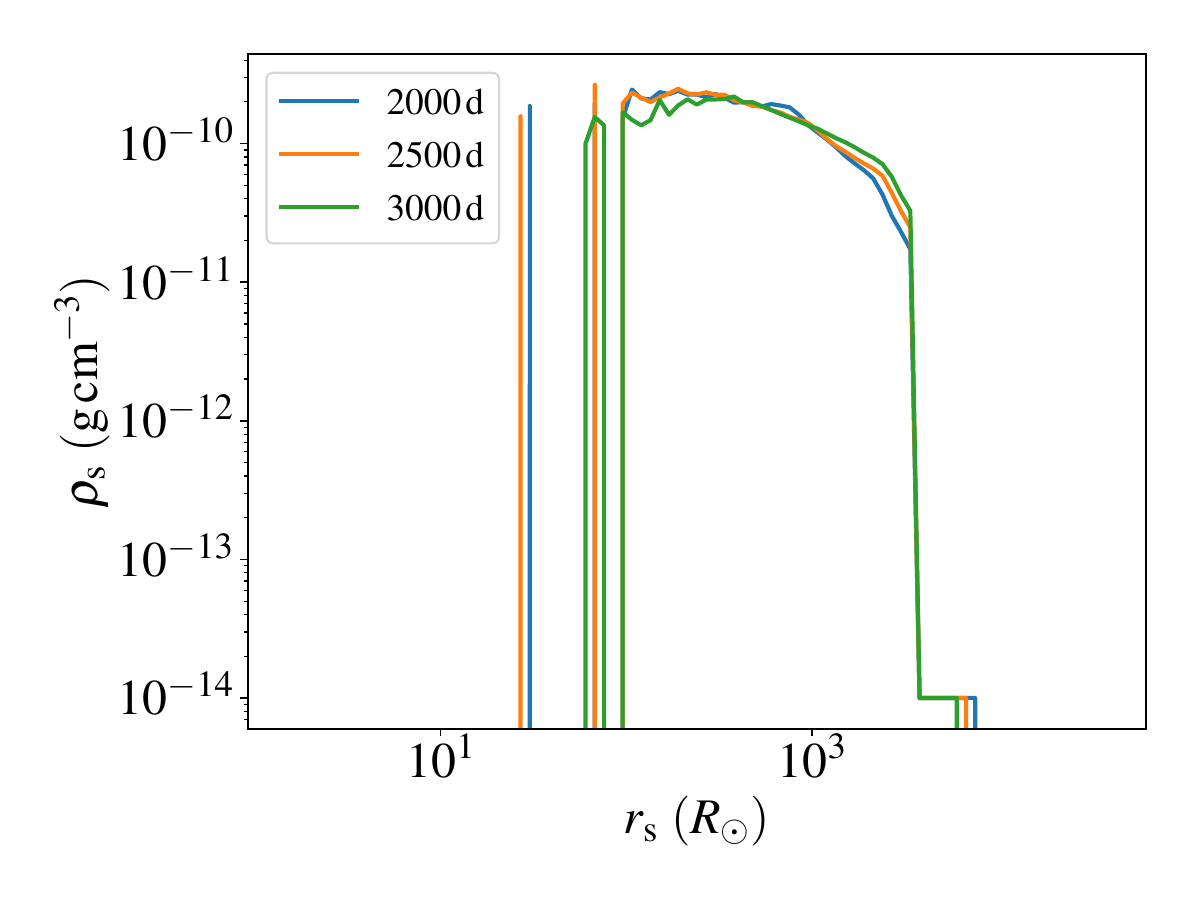}
    \caption{Radial velocity (top) and scaled density (bottom) as a function of scaled radius for different times ($2000$\,d, $2500$\,d and $3000$\,d). We also show a power-law fit (solid black line) for $v_{\rm r}\propto r_{\rm s}^{0.86}$ (top).}
    \label{fig:scaleddensity}
    \end{figure}

\begin{figure}
\includegraphics[width=0.47\textwidth]{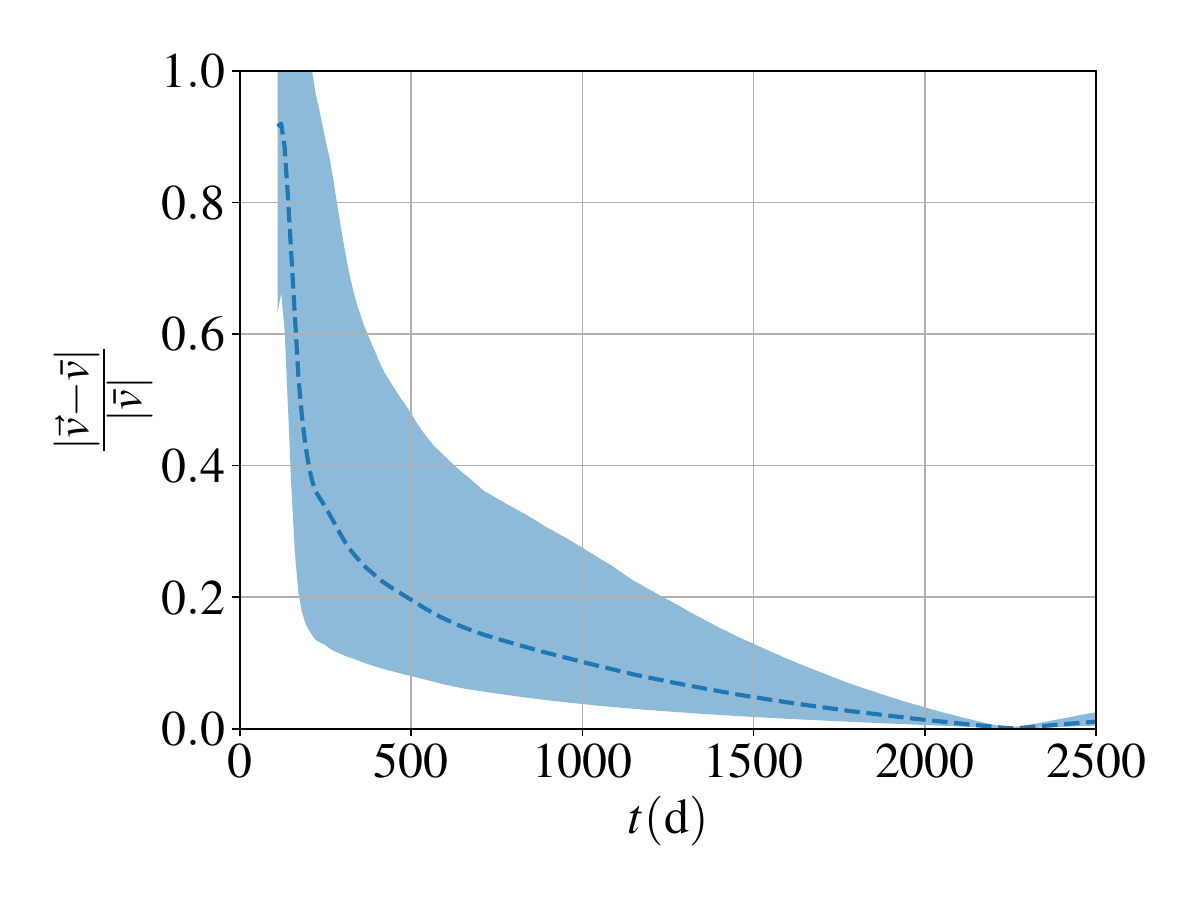}
\caption{The fractional change in the velocity of all particles compared to the final  average velocity as a function of time. The dashed line represents the mean  and the shaded area represents the standard deviation of the fractional change in the particle velocity.}
\label{fig:deltav}
\end{figure}

\begin{figure}
\includegraphics[width=0.47\textwidth]{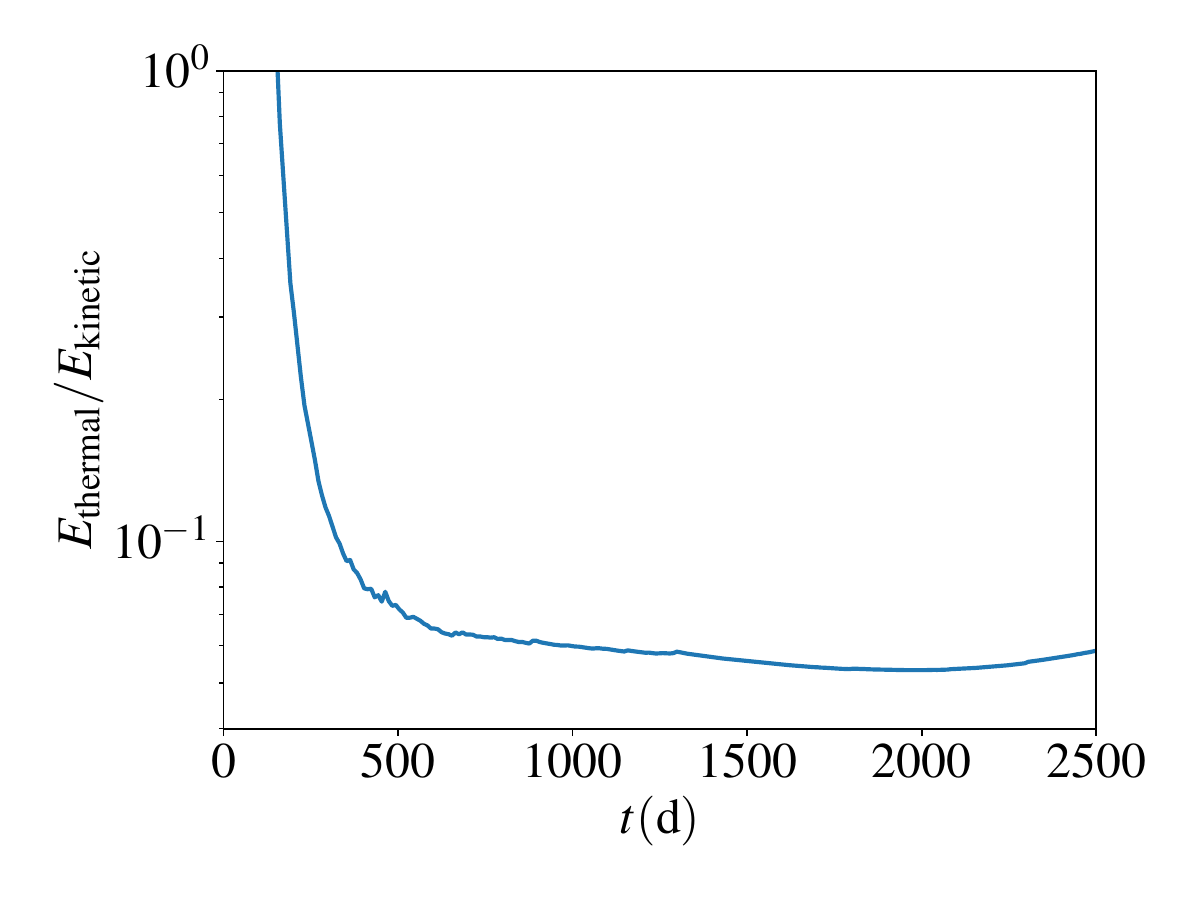}
\caption{Ratio of thermal energy to kinetic energy as a function of time.}
\label{fig:thermal}
\end{figure}

To show that our simulations follow the scaling defined by equations (\ref{eq:rs}) and (\ref{eq:scalings}), we select $t_{\rm i}\approx\tinitial\,$d and fit $t_0\approx \thstart\,$d so that the re-scaled radial velocities $v_{\rm r,s}$ match one another between a few $\times 10^2\,\rsun$ and a few $\times 10^3 \,\rsun$.  
We show this result in the top plot of Fig.~\ref{fig:scaleddensity}. 
As this plot shows, the velocities for $r_{\rm s}>$ a few $\times 10^2\,\rsun$ match each other for a number of different time steps.  
We also fit a power law between $6\times10^2\,\rsun$ and $5\times 10^3 \,\rsun$, and the resulting fit is
\be \label{eq:vr_300days}
\vr = 4.8 \left (\frac{r_{\rm s}}{100\, \rsun}\right )^{0.86}{\rm km\, s^{-1}}.
\ee
The radial power law exponent is about $1$, which is consistent with homologous expansion.  Thus we find that by $2000\,$d since the start of the simulation, homologous expansion is definitively reached.

We also plot the scaled density $\rho_{\rm s}$ in the bottom plot of Fig.~\ref{fig:scaleddensity}. Similar to the behaviour of  $v_{\rm r}$, we observe that the $\rho_{\rm s}$ match one another for a few different epochs when rescaled by $r_{\rm s}$. This is expected in the case of homologous expansion when the (scaled) density structure is frozen.



In Fig.~\ref{fig:deltav}, we show the fractional change in the absolute 3-D velocity of all fluid elements relative to their asymptotic (late-time) velocities approach zero as the envelope evolves. 
The fractional change is computed based on the late-time velocity of each fluid element, defined as the average of the velocity between $2000\,$d to $2500\,$d.
The dotted line in Fig.~\ref{fig:deltav} represents the mean of the fractional change in the particle velocity. 
The shaded region represents the standard deviation from the mean. 
Thus, the velocities of fluid elements do not change by more than 5 per cent either in the magnitude or direction after about $1500\,$d on average. 
This implies that the fluid elements are on ballistic trajectories. 

In addition, we plot the ratio of total thermal energy in the system to the total kinetic energy as a function of time in Fig.~\ref{fig:thermal}.   
The thermal energy is smaller than 5 per cent of the kinetic energy after 500\,d.  
The fact that this thermal energy does not continue to drop due to adiabatic expansion is because we use a temperature floor in our simulations to maintain numerical stability.  
In any case, thermal energy is a negligible fraction of the energy budget of the system.
 
\begin{figure}
    \includegraphics[width=0.47\textwidth]{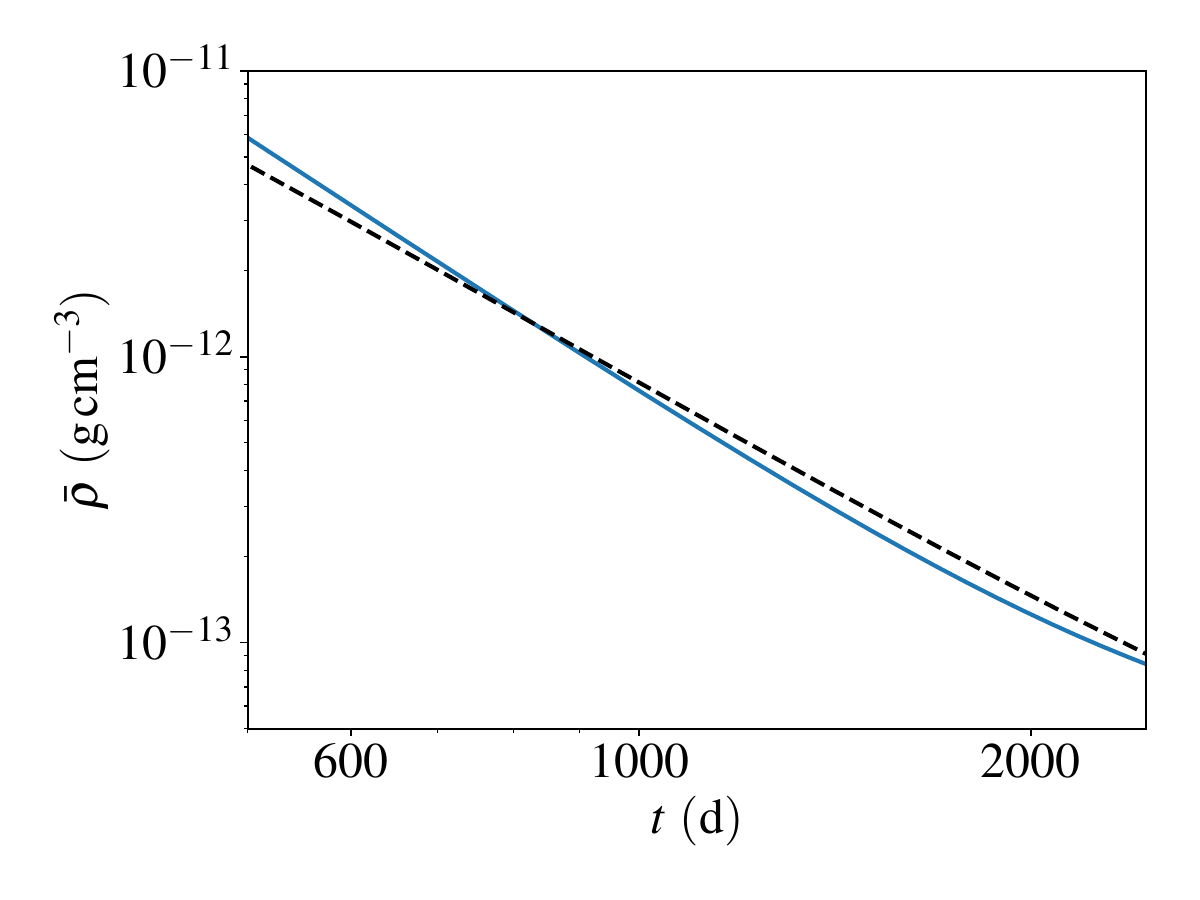}
    \caption{Mean density of the envelope as a function of $t$ (blue solid curve).  The black dotted curve is the $t^{-3}$ fit and is fitted for $t\in[500, 2000]$\,d .  Note that the average density profile follows a $t^{-3}$ power-law closely.} \label{fig:avgdensity}
    \end{figure}

Finally, for a homologously expanding system under adiabatic conditions, the average density scales like 
\be\label{eq:avgdensity_fit}
    \bar\rho \propto \thomo^{-3}.
\ee
This has also been previously demonstrated numerically by \citet{Iaconi:2019} for their SPH simulations.  
Here we confirm the same result by plotting the average density of unbound particles (solid line) in the envelope as a function of $t$ in Fig.~\ref{fig:avgdensity}. We also plot a $t^{-3}$ power law fit (dashed line) that is fitted for $t\in[500, 2000]$\,d.  The average density from our simulation follows the $t^{-3}$ power law.   



\section{1-D Model} \label{sec:1d}

\begin{figure}
    \centering
    \includegraphics[width=0.47\textwidth]{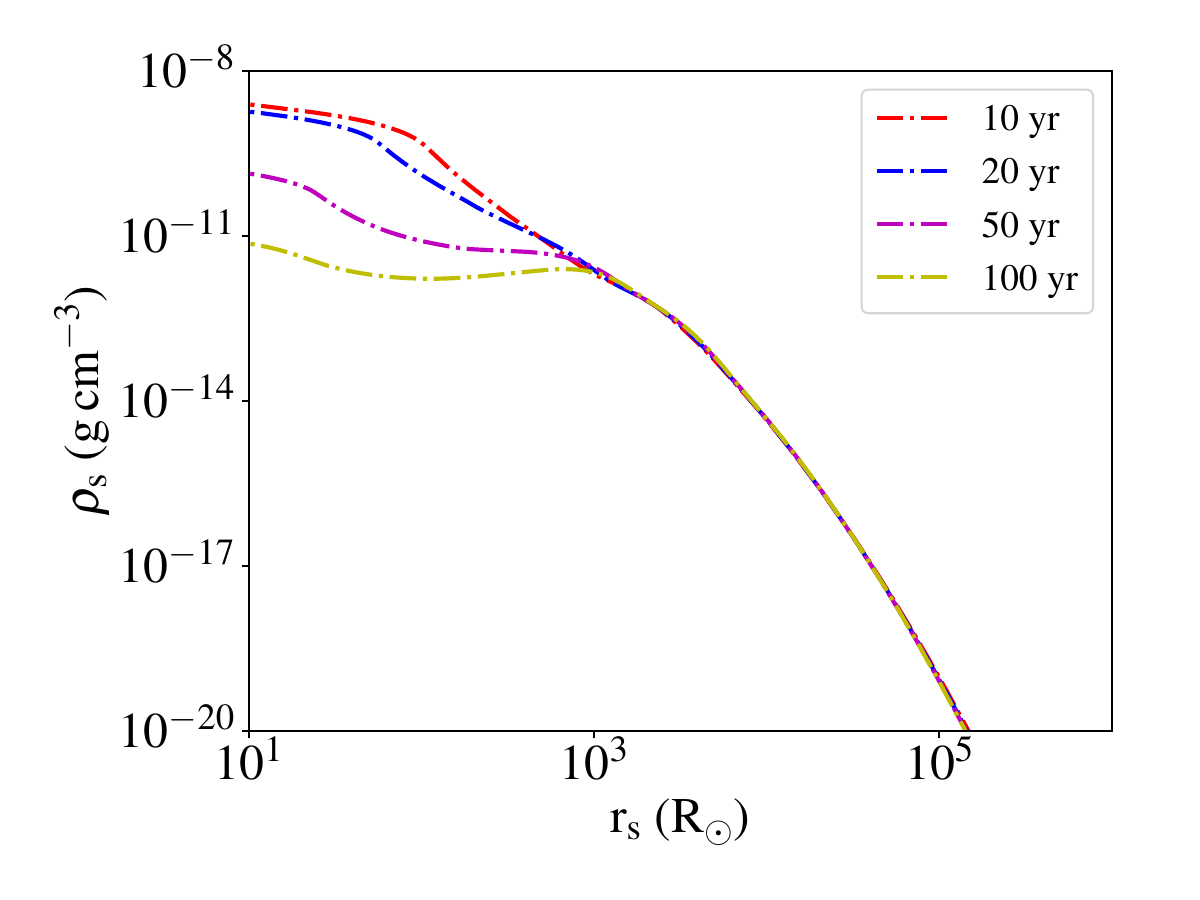}
    \caption{Scaled density, $\rho_{\rm s}$ as a function of scaled radius, $r_{\rm s}$ from 1-D simulations, for $10$ (red), $20$ (blue), $50$ (magenta), and $100$ (yellow) yr and a heating parameter of $\lambda=1$. }
    \label{fig:rho1D}
\end{figure}

\begin{figure}
    \centering
    \includegraphics[width=0.47\textwidth]{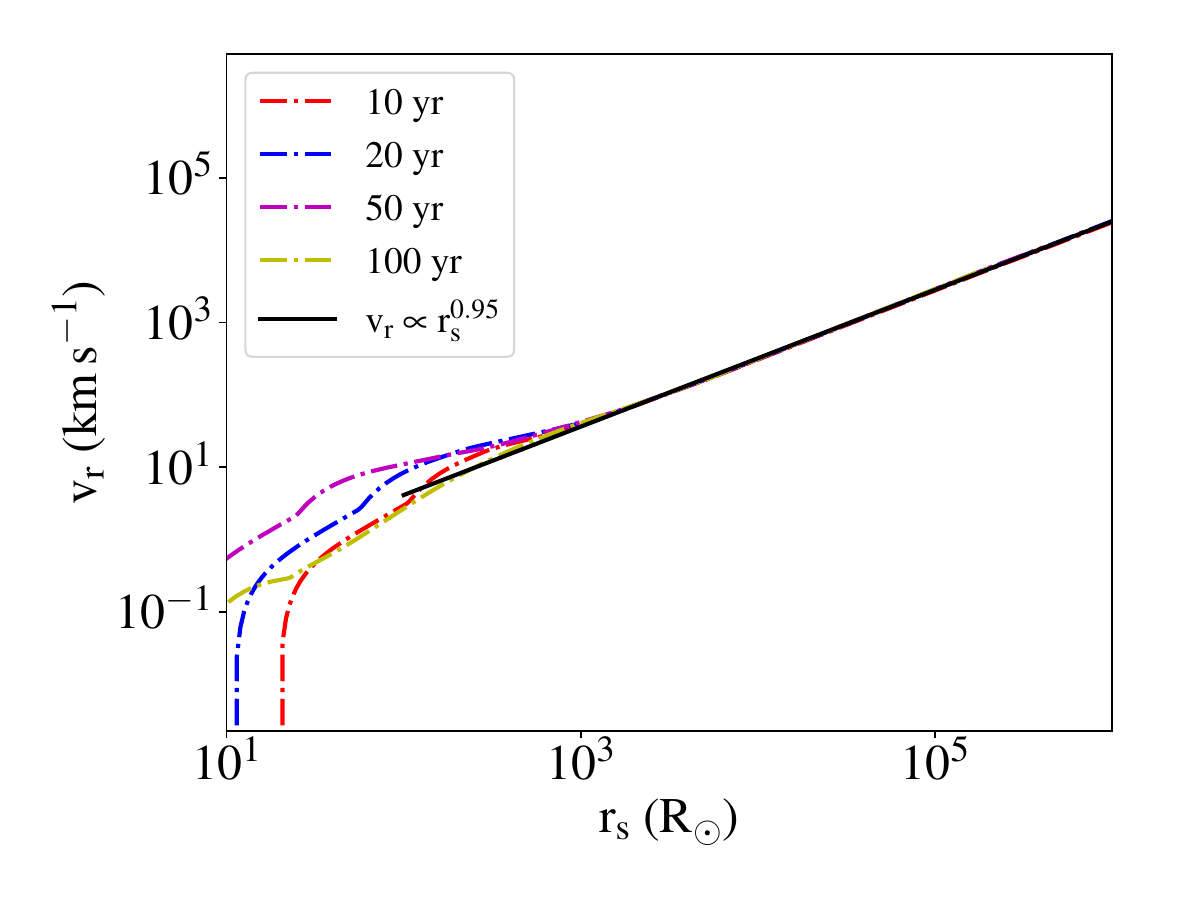}
    \caption{Radial velocity, $\vr$, as a function of scaled radius from 1-D simulations. The times and $\lambda$ are the same as in Fig.\ref{fig:rho1D}. The solid black line is the best fit of $\vr \propto r_{\rm s}^{0.95}$.     \label{fig:vel1D}}
\end{figure}

Motivated by the results of the previous section, we now study a simplified 1-D spherically-symmetric model of the ejected envelope.  
We have developed a 1-D finite-volume spherically-symmetric hydrodynamics code in Python that uses an HLLE Riemann solver~\citep{HLLE1983, HLLE1988} with piecewise-constant (first-order) reconstruction to study this ejected envelope\footnote{This 1-D code will be shared on reasonable request to the corresponding author.}. 
Our models consist of 350 grid points that are logarithmically spaced, starting from $r=3\,\rsun$ to $r=3\times 10^7\,\rsun$, giving 50 grid points per decade. 
We set the origin to the centre of mass of the binary and use free (von Neumann) boundary conditions on the inner and outer boundaries. 
While the discussion of the hydrodynamic equations can be widely found in the literature, we will briefly recap them here for completeness.  
These equations can be written in compact notation by introducing a state vector $\boldsymbol{\mathcal{U}}=(\rho,\rho \vr,\rho e)$
\be\label{eq:conservative}
    \frac{\partial\boldsymbol{\mathcal{U}}}{\partial t}+  \frac{1}{r^2}\cdot\frac{\partial\boldsymbol{\mathcal{F}}}{\partial r}=\boldsymbol{\mathcal{S}},
\ee
where $\boldsymbol{\mathcal{F}}=(r^2\rho \vr,r^2(\rho \vr \vr + P),r^2(\rho e+P) \vr)$ is the flux function,  
$\boldsymbol{\mathcal{S}}=(0,-\rho\frac{GM_{\rm r}}{r^2} ,-\rho \vr \frac{GM_{\rm r}}{r^2}+ \mathcal{S}_{\rm h})$ 
is the source function, $e$ is the specific energy, $G$ is the gravitational constant and $M_{\rm r}$ accounts for the mass of the central binary and the envelope enclosed within radius $r$. 
The extra term, $\mathcal{S}_{\rm h}$, is added to study the effect of the heating supplied to the envelope from the central binary. We discuss this below.
For the initial conditions of the 1-D model, we take the fitted results from the 3-D numerical simulation at $t=800$\,d. 


The 1-D model computes $1000$ outputs over $100$\,yr, each one separated by $0.1$\,yr. 
We present the results in Figs.~\ref{fig:rho1D} and \ref{fig:vel1D} for $10$, $20$, $50$, and $100$ yr. 
These times are relative to the start of the 3-D simulation so that the same time between the two simulations can be directly compared. The scaled radius and density follow equations (\ref{eq:rs}) and (\ref{eq:rhos}), respectively. 
We calculate the best fit of the linear part to get the power-law relation between $\vr$ and $r_{\rm s}$. 
We fit each time step separately and average them to produce a best fit of $\vr \propto r_{\rm s}^{0.95}$.

We note that the 1-D model is intentionally not constrained to adhere to the homologous expansion model, though we did use initial conditions that correspond to the beginning of the homologous phase.
One feature that is observed in these 1-D models, but not in the full 3-D models, is evident in  Fig.~\ref{fig:vel1D}.  
Here at a radius below about a few\,$\times 10\,\rsun$, the radial velocity becomes negative. 
This manifests as the vertical rise in Fig.~\ref{fig:vel1D}. 
This is due to the gravitational potential from the inner binary being much larger than the total energy of the envelope in this region.  
However, these negative radial velocities are not seen in full 3-D models though the data is quite noisy in this region (Fig.~\ref{fig:scaleddensity}). 

This difference may be attributed to the periodic forcing of the orbiting binary on the gas in this inner region.  To examine this, we develop a simple model of binary heating for this 1-D model.
The heating from the central binary can be thought of as a periodic forcing from the binary driving a damped simple harmonic oscillator with a frequency equal to the epicyclic frequency, $\kappa = \Omega$, where $\Omega$ is the Keplerian orbital frequency.  A discussion of this heating term is in Appendix \ref{appendix}, but the resulting parameterized heating term is
\be\label{eq:heating_source}
\mathcal{S}_{\rm h} =  \lambda\Omega\frac{GM_{\rm bin}\rho}{r} \left(\frac{a_{\rm bin}}{r}\right)^2,
\ee
where $\lambda$ is a free parameter, $M_{\rm bin}$ is the binary mass and $a_{\rm bin}$ is the binary separation. Here, we set $M_{\rm bin} = 1.36\,\msun$, and $a_{\rm bin} = 5\,\rsun$.

\begin{figure}
    \centering
    \includegraphics[width=0.49\textwidth]{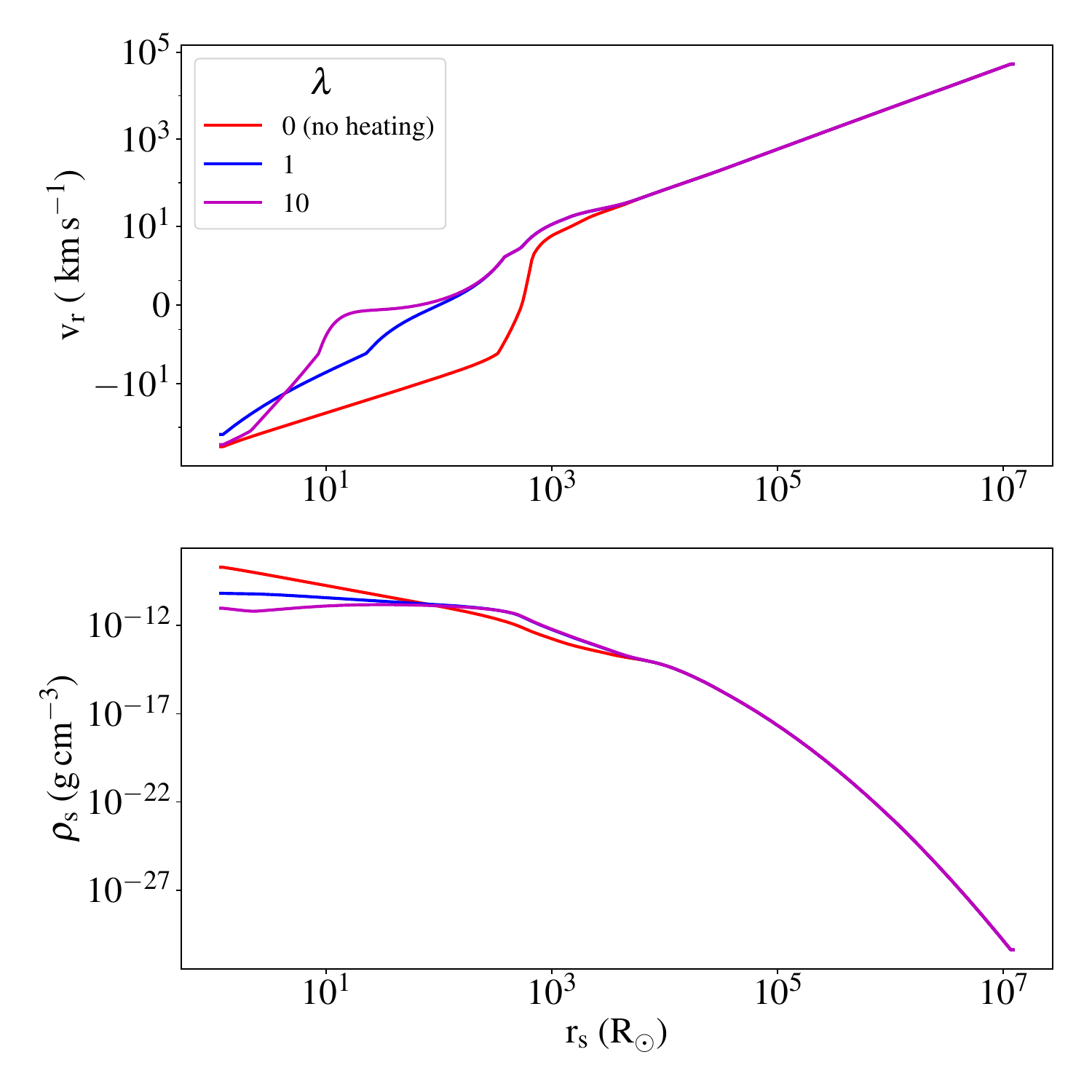}
    \caption{Scaled density ($\rho_{\rm s}$) and radial velocity ($\vr$) as a function of scaled radius ($ r_{\rm s}$), for 10 yr and different heating parameters $\lambda$.}
    \label{fig:lambda_comparison}
\end{figure}

Fig.~\ref{fig:lambda_comparison} shows $v_{\rm r}$ and $\rho_{\rm s}$ at $10$\,yr for different heating parameters $\lambda$. 
The heating term $\mathcal{S}_{\rm h}$ impacts the inner envelope and has little impact in the outer regions ($r_{\rm s}>10^4\,\rsun$).  
For the case of no heating ($\lambda=0$, red line), the inner regions follow a free-fall inflow solution.  
For larger heating rates, this inflow is suppressed but not completely eliminated, though the infall region occurs at substantially smaller radii than is effectively probed by our 3-D simulations. 
In either case, the heating rate makes no impact on either the density or velocity profile at large radii.  
We thus conclude that binary heating or any other late-time heating has little effect on the ultimate expansion and ejection of the envelope, but is required to prevent the infall of the innermost envelope.  
In the 3-D simulations, angular momentum or turbulence of the gas close to the binary may play a similar role, but this is not well modelled in the 1-D simulation. 
In any case, it is evident that there is extra physics that is not entirely accounted for in the 1-D model that leads to homologous expansion (at smaller radii) in the 3-D model.

\section{Discussion}\label{sec:discussion}
In this work, we study the long-timescale evolution of CEE and the homologous nature of envelope expansion. We simulate a common-envelope event using \changaMM~for a $2\,\msun$ RGB and a $1\,\msun$ MS star binary system. We show that nearly all gas from the red giant is unbound in \unbfull\,d using an adiabatic equation of state and relying only on the release of orbital energy.

This is in agreement with and in contrast to other work.  For instance, \cite{Chamandy:2020} evolved a binary system of an AGB + white dwarf or MS star system through 20 orbits using an adiabatic equation of state. 
In agreement with our findings, they show that the envelope unbinds at a constant rate and would become unbound in less than $10$\,yr.

On the other hand, a number of others  suggest that additional energy injection is necessary.
\cite{Ondratschek:2022} studied a binary system consisting of an AGB primary similar to \cite{Sand:2020} with a white dwarf or a MS star such that the mass ratio is 0.25 using the OPAL equation of state \citep{2002ApJ...576.1064R}.
They find a complete envelope ejection in about $1000$\,d when considering thermal and ionization energy along with mechanical energy, and in about $3400$\,d when ignoring the thermal and ionization energy. 
\cite{Sand:2020} studied the fraction of unbound mass in two different simulations, one using the ideal gas equation of state and the other using the OPAL equation of state \citep{2002ApJ...576.1064R} for a binary system with an AGB primary and a white dwarf or an MS companion star such that the mass ratio is 0.5.  
In the case of the ideal gas equation of state, only $20$ per cent of the mass becomes unbound and the rate of mass ejection is slower if the internal energy is ignored.  
In their OPAL runs, $80$~per~cent of the mass is unbound in about $2500$\,d considering only the mechanical energy and $100$~per~cent is ejected in about $1000$\,d considering mechanical energy along with thermal and recombination energy. 
Finally, \cite{Iaconi:2019} find that the envelope of a binary system with  $0.88\,\msun$ RGB and a $0.6\,\msun$ MS companion star is completely ejected in about $500$\,d when considering mechanical and recombination energies. 

One crucial difference between our work and that of \citet{Sand:2020} and \citet{Ondratschek:2022}  is that the cores of the stars end up in a much tighter binary in our case.  
In particular, \citet{Sand:2020} starts the binary at $236~\rsun$ but ends up at $41~\rsun$.  
In our case, we start at around $50~\rsun$, but end up at $5~\rsun$.  
Hence, our orbit shrinks by a factor of nearly $10$ whereas \citet{Sand:2020}'s orbit shrinks by a factor of about $5$.  
The corresponding relative gravitational energy release is hence a factor of 2 greater in our case.


In addition, we also show that the envelope reaches homologous expansion starting around a few hundred days ($\thstart$\,d). 
In comparison, \cite{Iaconi:2019} showed that it takes about $5000$\,d for the bulk of the unbound gas to become homologously expanding, even though the external layers of the envelope become homologous as soon as they are ejected. 
This difference may be due to the analysis methodology.  
Namely, \citet{Iaconi:2019} traced the ballistic trajectories of SPH particles whereas we looked at the velocities of fluid elements and fit the velocities and associated radii to a $\thomo = 0$ point.
We are also simulating these events at substantially higher resolution through a combination of greater particle count and the use of a moving-mesh methodology. 

Inspired by the homologous expansion in our 3-D simulation, we also study a 1-D model for $t>\tinitial$\,d. 
Unsurprisingly, we find that initializing the 1-D simulation with the spherical approximation of the 3-D simulation data produces homologous expansion in the bulk of the envelope.  However, the inner regions require some additional physics not modelled in the 1-D simulation to preclude fallback and, hence, the breaking of homology.  Here, we attribute this additional physics to heating from the periodic forcing of the binary but note that turbulence or angular momentum may play the same role.

The fact that simulated common-envelope events follow both a (roughly) spherical and homologous expansion approximation is likely useful for their theoretical and observational studies.  First, CEE need not be numerically simulated for extremely long timescales.  Instead, they only need to be simulated to the point where they begin homologous expansion, which occurs on a timescale of years as opposed to decades.  This will result in significant computational savings and an associated expansion in the parameter space that can be explored.  

Secondly, the fact that they obey both the spherical and  homologous expansion approximations implies that radiation transfer codes that are used to calculate supernova light curves and spectra can be adapted to compute light curves and spectra from CEE events. 
Homologous expansion kinematic models are widely accepted as a good first-order approximation to model radiation transfer in supernova explosions \citep{Ropke:2005,2014MNRAS.440..387K,Liu:2018}. 
These radiation transfer codes typically assume a spherical and homologous expansion profile for the ejecta to greatly speed up the calculation.  
This fact has already been utilized by \citet{Kaminski:2018} to model the CO emission in V4332 Sgr as a homologously expanding bipolar flow. 
Similarly \cite{Kaminski:2020} show that the observed properties of the molecular remnant of Nova 1670 (CK Vulpeculae) are reproduced by assuming linear velocity fields.

Finally, we note that we have neglected effects, such as magnetic fields \citep{Ondratschek:2022} and jets \citep[see for instance][]{2021RAA....21...90S} that could cause the outflow to become more non-spherical at late times.
The effect of jets in CEE is still unclear, as \citet{Zou:2022} found that jets are quickly choked within the envelope. 
On the other hand, \citet{Ondratschek:2022} showed that late-time jet-like outflows produce the bipolar morphology seen in many planetary nebular systems.  These effects will be a topic of future work.

\section{Conclusions}\label{sec:conclusions}
In this work, we have analyzed the nature of expansion and the timescale of complete envelope ejection in common-envelope evolution. 
We used the moving-mesh hydrodynamic solver \changaMM~to perform a long-timescale simulation of a  CE system involving a $2\,\msun$ red giant and a $1\,\msun$ main sequence star. 
We let the system evolve for around $13$\,yr. 
Starting at an orbital radius of $52\,\rsun$, the binary plateaus to an orbital radius of $5\,\rsun$ in $200$\,d. 
We observe that nearly all envelope material is unbound after \unbfull\,d, and $80$ per cent is unbound in \unbeighty\,d. 
This timescale is similar to that found by others who also studied envelope ejection in low-mass binary systems.  
However, we find that there is no need for additional energy injection from recombination.

Motivated by previous work by \citet{Iaconi:2019}, we also show that the envelope enters a phase of homologous expansion after \thstart\,d. 
This is likely important for theoretical and observational work on CEE.
First,  one can have significant computational savings by doing numerical simulations up to the homologous start time and using the homologous expansion model afterwards. 
Secondly, the radiative transfer codes used for finding light curves and spectra for supernovae can be adapted for use in CEE simulations.

Finally, we study the homologous expansion model in 1-D simulations using the power-law fits of the homologous phase as initial conditions. 
From this study, we found that periodic heating from the binary star at late times can affect the inner regions of the envelope but does not impact the homologous expansion. \strikeout{Also, from the analytic analysis, we showed that homologous expansion is naturally expected for a power-law analysis of the 1-D spherically symmetric hydrodynamics equations.}

\section*{Data Availability}

The data underlying this article will be shared on reasonable request to the corresponding author.

\section*{Acknowledgements}
VV acknowledges support from the NSF through grants PHY-1912649 and PHY-2207728. 
SvB, LP, and PC acknowledge support from the NASA ATP program through NASA grant NNH17ZDA001N-ATP, the NSF through grant AST-2108269, and the UWM Research Assistance Fund. 
LP also acknowledges support from the UWM R1 Distinguished Dissertator Award, NSF through grant PHY-1748958 and a grant from the Simons Foundation (216179, LB). 
\changaMM~simulations were completed on the Mortimer HPC System at UWM, which was funded by the NSF Campus Cyberinfrastructure Award OAC-2126229 and UWM.
We also use the yt software platform for the analysis of the data and generation of plots in this work \citep{yt}.

\label{lastpage}
\bibliographystyle{mnras}
\bibliography{bib}
\appendix
\section{Simple Harmonic oscillator model for heating from central binary}
\label{appendix}

For a fluid element in a circumbinary orbit around two point masses, we model the periodic forcing from the inner binary as a forced simple harmonic oscillator with a natural resonant frequency of $\kappa$, which is the epicyclic frequency.  
The periodic forcing term has a frequency of $m(\Omega_{\rm b} - \Omega)$, where $m$ is some natural number, $\Omega_{\rm b}$ is the orbital frequency of the binary and $\Omega$ is the orbital frequency of the fluid element.

A general driven damped harmonic oscillator equation is then

\be
    \frac{d^2\delta r}{dt^2}+\Gamma\frac{d\delta r}{dt} + \kappa^2 \delta r = f_0\cos{m(\Omega_{\rm b} - \Omega) t},
\ee
where $\Gamma$ is a damping term that arises from fluid dissipation and $f_0$ is the overall forcing amplitude.  Letting $\omega = m(\Omega_{\rm b} - \Omega)$, the complex form of this equation is
\be
    \frac{d^2\delta r}{dt^2}+\Gamma\frac{d\delta r}{dt} + \kappa^2 \delta r = \frac{F_0}{m}\exp(i\omega t),
\ee
Starting with an ansatz, $\delta=\mathcal{A}\exp{i\omega t}$, we can show
\be
     \mathcal{A} = f_0 \frac {\kappa^2 - \omega^2 - i\omega\Gamma}{(\kappa^2-\omega^2)^2+(\Gamma\omega)^2},
\ee
where we can write 
\be
     \mathcal{A} =A\exp{i\delta}
\ee
where 
\be
     A=\frac{f_0}{\sqrt{(\kappa^2-\omega^2)^2+(\Gamma\omega)^2}} \\
     \delta = \tan^{-1}\left(-\frac{\omega\Gamma}{\kappa^2-\omega^2}\right)
\ee
The specific energy is then
\be
    \epsilon = \frac{1}{2}\dot{\delta r}^2 + \frac{1}{2} \kappa^2{\delta r}^2
\ee
The maximum amplitude, $A$, occurs when $\omega=\kappa$ and is:
\be
    A=\frac{f_0}{\Gamma\kappa},\qquad\textrm{and}\qquad \delta = \frac {\pi} 2,
\ee
and the maximum specific energy is:
\be
    \epsilon_{\rm max} = \frac{1}{2}\frac{(f_0)^2}{\Gamma^2}
\ee
The specific energy dissipation is thus,
\be
    \dot\epsilon = 2\epsilon_{\rm max}\Gamma = \frac{f_0^2}{\Gamma}
\ee

For our case, we can model the forcing from the inner binary as a tidal forcing
\be
f_0 = \beta \frac {G M_{\rm bin} a_{\rm bin}}{r^3},
\ee
where $\beta$ is some constant, $M_{\rm bin}$ is the mass of the binary, $r$ is the orbital radius of the fluid element, and $a_{\rm bin}$ is the separation of the two stars. Likewise, we can set the damping rate as some fraction of the epicyclic frequency $\Gamma = \alpha\kappa$. The volumetric heating rate is then
\be
    \mathcal{S}_h =  \frac{2\beta^2}{\alpha}\frac{\kappa GM_{\rm bin}\rho}{r} \left(\frac{a_{\rm bin}}{r}\right)^2.
\ee
Here we define the relevant constant as $\lambda \equiv 2\beta^2/\alpha$.  Combining these terms together and setting $\kappa = \Omega$ in Keplerian potentials, we arrive at equation (\ref{eq:heating_source}).

\end{document}